# Research
## A SCIENCE PARTNER JOURNAL

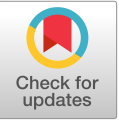

# Securing Data in Multimode Fibers by Exploiting Mode-Dependent Light Propagation Effects

**Stefan Rothe[1]\*, Karl-Ludwig Besser[2], David Krause[1], Robert Kuschmierz[1], Nektarios Koukourakis[1], Eduard Jorswieck[2]\*, and Jürgen W. Czarske[1,3]\***

[1]TU Dresden, Faculty of Electrical and Computer Engineering, Laboratory of Measurement and Sensor System Technique, 01062 Dresden, Germany. [2]TU Braunschweig, Faculty of Electrical Engineering, Information Technology, Physics, Institute for Communications Technology, 38106 Braunschweig, Germany. [3]TU Dresden, Faculty of Physics, 01062 Dresden, Germany.

*Address correspondence to: stefan.rothe@tu-dresden.de (S.R.); e.jorswieck@tu-bs.de (E.J.); juergen.czarske@tu-dresden.de (J.W.C.)

Multimode fibers hold great promise to advance data rates in optical communications but come with the challenge to compensate for modal crosstalk and mode-dependent losses, resulting in strong distortions. The holographic measurement of the transmission matrix enables not only correcting distortions but also harnessing these effects for creating a confidential data connection between legitimate communication parties, Alice and Bob. The feasibility of this physical-layer-security-based approach is demonstrated experimentally for the first time on a multimode fiber link to which the eavesdropper Eve is physically coupled. Once the proper structured light field is launched at Alice's side, the message can be delivered to Bob, and, simultaneously, the decipherment for an illegitimate wiretapper Eve is destroyed. Within a real communication scenario, we implement wiretap codes and demonstrate confidentiality by quantifying the level of secrecy. Compared to an uncoded data transmission, the amount of securely exchanged data is enhanced by a factor of 538. The complex light transportation phenomena that have long been considered limiting and have restricted the widespread use of multimode fiber are exploited for opening new perspectives on information security in spatial multiplexing communication systems.

## Introduction

The channel capacities in optical networks can be increased by several orders of magnitude using space-division multiplexing (SDM) in complement to established methods [1–3]. Contrary to single-mode fibers (SMFs) that primarily provide wavelength or polarization multiplexing, spatial degrees of freedom are accessible with multimode fiber (MMF) or multicore fibers. These advancements are the result of intense research over the past decade, after Chraplyvy predicted the emerging capacity crunch and proposed space as a new essential degree of freedom in communications technology [4]. The requirement for sophisticated SDM solutions is particularly important, as the global exchange of data is increasing exponentially [5] and the number of connected devices in the evolution of Internet-of-Things applications in the 5G era and beyond increases dramatically [6,7]. This development requires not only higher data rates but also improved security in terms of confidentiality or privacy with low complexity [8] that can resist a quantum computer.

However, fiber-based data links are prone to various security risks [9–11]. In general, fiber cables, especially SMF cables, are susceptible to evanescent attacks at intentionally inserted bending points [12,13]. At positions with high optical power, for example, at outputs of amplifiers, eavesdroppers can remain undetected. In such cases, it is of utmost importance that messages encoded in tapped signals cannot be reconstructed by unauthorized receivers, which is why additional arrangements need to be made for enabling information security.

Within the past decades, an enormous amount of research has been done in the field of quantum communication [14,15]. Compared to all alternatives, quantum key distribution (QKD) shows the strong benefit that the achievable security is underpinned by physical laws. Hence, information-theoretic security can be realized. Although hybrid QKD links consisting of both free-space satellite-to-ground and fiber links with a large number of nodes can already be built in practical environments, the achievable key rate is still in the kilohertz range [16]. This can be attributed to severe challenges that arise by generating single photons at high rates in defined time intervals, e.g., by quantum dots [17,18]. In addition, each component used in a QKD system induces undesired photon losses. In particular, this poses enormous challenges for the integratability into multimode SDM networks [19]. The benefit of quantum security is also its disadvantage regarding long-haul transmission system, since the realization of quantum repeaters requires fundamental paradigm shifts in quantum mechanics [20]. Although QKD systems offer information-theoretic security, it is desirable to consider possible alternatives for information security, especially for SDM networks.

Conventionally, sensitive data are secured at high communication layers either through symmetric key or asymmetric key cryptography. However, a common flaw of symmetric cryptographic approaches is the need for key exchange. This drawback is addressed in asymmetric approaches also known as public-key cryptography, where key exchange is omitted. Nevertheless, they are considered weak compared to quantum computers that, in the future, could efficiently solve prime factorization problems in polynomial time [21]. In Wyner's seminal paper [22], it was shown that there exists a class of channel codes called wiretap codes, which allow a reliable and confidential communication exploiting actual physical channel properties without a key exchange between two users. Wyner demonstrated that, within practical implementations, information-theoretic security is achievable using wiretap codes, which can provide the same security level as reached by QKD. This work was related to noisy channels and has opened the research field of physical layer security (PLS). The approach is to exploit the properties of the physical channel between legitimate communication parties, transmitter (Alice) and receiver (Bob), to make decipherment of tapped data as difficult as possible for a non-legitimate receiver (Eve).

Light propagation through disordered MMF channels is characterized by several complex phenomena, such as modal crosstalk (XT) and mode-dependent loss (MDL). Although these effects appear unpredictable in individual scenarios, they can be measured and compensated. This makes the MMF a suitable candidate for implementing PLS using spatial modes as the physical carrier of information. Once the MMF channel properties are known to Alice by transmission matrix (TM) measurements [23–25], she is able to launch a suitable wavefront with a spatial light modulator (SLM) equalizing the channel to Bob that is referred to as channel diagonalization in the following. Since there is a mismatch between the channels Alice–Bob and Alice–Eve, this diagonalization does not apply to Eve, who has to invert her channel, for example. Alice and Bob can take advantage of this asymmetry by generating a proper wiretap code.

Here, we present the first experimental implementation of confidential data transmission using PLS in an MMF-based communication scheme. The approach introduced is implemented to real MMF to which Eve is physically coupled. In a calibration step using sequential mode launching by means of an SLM and digital holography, the TM of the MMF under test is measured. In our investigations, we employ channel diagonalization based on singular value decomposition (SVD) that is applied to the TM measured between Alice and Bob. This creates a data connection through which we perform data exchange. The use of wiretap codes enables to harness the asymmetry between Bob and Eve securing the data link. We create wiretap codes based on the measured TMs and apply them experimentally to the MMF showcasing the power of PLS. Using Monte Carlo (MC) simulations and a suitable model tailored to the MMF channel, we are able to analyze the measured TMs between Alice and both Bob and Eve determining the achievable number of bits that can be exchanged securely, i.e., secrecy rate $R_S$. Contrary to conventional public key algorithms based on prime factorization, e.g., asymmetric key cryptography, instantaneous physical parameters are exchanged within our approach, which cannot be predicted by a quantum computer [26] contributing to postquantum cryptography.

## Results

### PLS in MMFs

Once coherent light is launched at the input of MMF, it emerges as an randomly appearing light field at the output, known as a speckle pattern. Inherent phenomena such as XT [27] and MDL [28] are the source of this effect and are strongly dependent toward the direction of light propagation. Thus, at any position $z_E$ between the two opposite MMF facets representing legitimate communication parties, an observer receives a different light field, as shown in Fig. 1. With this asymmetry, it is possible to take advantage of the complex light propagation through MMF, making it a suitable candidate for PLS, which has been introduced in preliminary work [29].

As a consequence of asymmetric disorder induced by XT and MDL, the measured channel matrix of the legitimated communication parties Alice and Bob is different to the channel between Alice and Eve, who gains optical access illegitimately at any position between Alice and Bob (excluding Alice and Bob's position). In Fig. 2A and B, measured TMs of these two channels from a 10-m-long MMF are shown. Using the measured TM, Alice and Bob can calibrate their channel and compensate for XT. For example, Alice is able to transmit modes to Bob by performing appropriate wavefront shaping on her side. However, this optical prescrambling does not diagonalize the channel to Eve, whose TM is different (see Fig. 2C). Eve must, for instance, invert her TM or perform any other operation to compensate for XT. However, the MDLs of an MMF channel are characterized by the distribution of the singular values in the TM, where a large difference between the highest and lowest values can be observed (see Fig. 2D, more than 60 dB difference). Hence, singularities can be observed in the inverse (see Fig. 2E). This has dramatic consequences for Eve, because noise becomes amplified during inversion. For Alice and Bob, the situation is different. Since they both measure their TM and Alice wants to diagonalize the channel to Bob, they can resort to alternative approaches. For example, channel diagonalization based on SVD is particularly well suited for this purpose. With SVD, two unitary matrices, $U$ and $V^H$, respectively, can be calculated (see Fig. 2F and H) that are used for both prescrambling and retrieving the exchanged information transmitted through the MMF. Compared to inversion, the dynamic range between highest and lowest magnitudes (see color bars) of $U$ is reduced substantially, which facilitates the optical implementation using the SLM. This method is described in the following section.

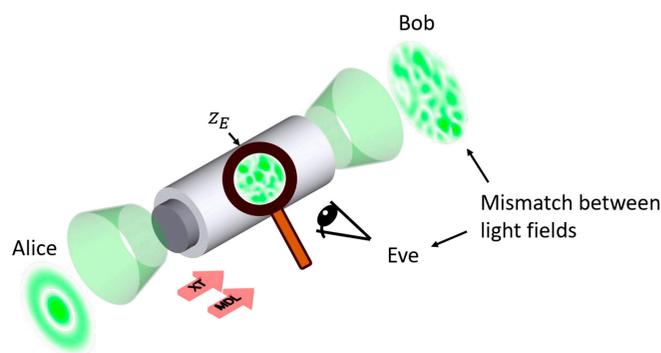

**Fig. 1.** Light propagation through MMF. When coherent light with arbitrary distribution is launched into MMF, XT and MDL induce a transformation yielding to a speckle pattern at the output. An observer at any position $z_E$ between input and output sees a different light field.

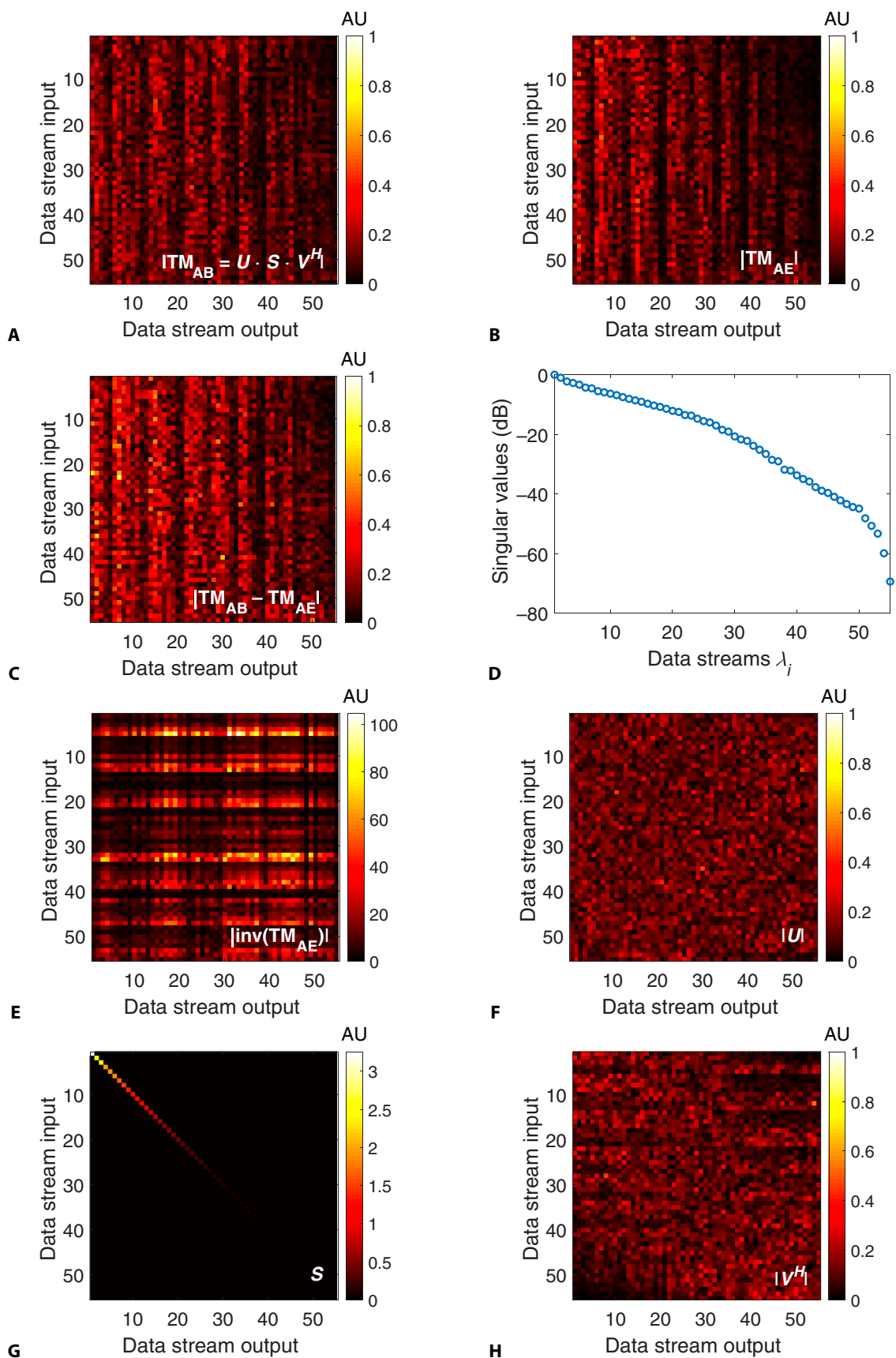

**Fig. 2.** TM occupancies according to 10 m 55 × 55 MMF channels with highly developed XT (amplitude values are shown). (A and B) TMs between Alice and both Bob and Eve. (C) Difference. (D) MDL distribution derived from $TM_{AE}$. (E) Inverse of $TM_{AE}$. Note that due to low singular values of $TM_{AE}$, singularities can be observed in the inverse. Contrary, the 3 matrices $U$, $S$, and $V^H$, which are determined from applying SVD to $TM_{AB}$ shown in (F) to (H), are well balanced regarding their dynamic range. AU, amplitudes in arbitrary units.

Once Alice and Bob have diagonalized their channel, they can exchange data. In the simplest case, they select the data streams according to high singular values exhibiting a strong information flow. If they want to enable confidentiality for their channel, then they can generate a wiretap code, for example. It is well known that wiretap codes are one way to achieve information-theoretic security [30]. Recall that we are considering the definition of strong secrecy from information-theoretic security [30, chapter 3.3]. The basic idea behind wiretap codes is to use a stochastic encoder where each message is randomly mapped onto one of multiple codewords. When designing the wiretap code, differences in the channels to Bob and Eve are exploited such that Bob is able to reliably decode the exact message, while Eve can only decode a set of possible codewords. Because of the stochastic encoding, this does not give her any information about the secure message since they are all equally likely. Even a quantum computer could not help Eve since she lacks the information. A detailed description can be found in the Information-theoretic security with wiretap codes section.

Although wiretap codes are well established for wireless networks, they are not straightforwardly transferable to MMF channels. The challenge is that fundamental signal transmission specifications are changed by switching the communication environment. One major difference is the choice of appropriate power constraints. While there is typically an average power constraint in wireless systems [31], instantaneous amplitude constraints need to be applied in MMF [32]. The motivation behind this stems from the optical setup where a laser is employed at the transmitter and a camera at the receiver. First, the laser on the transmitter side has a finite power output, and, second, the camera is calibrated for a certain maximum power to avoid nonlinear fiber effects. Besides power constraints, the nature of the transmission medium itself is fundamentally different in fiber-based communication. Thus, the comprehensive development of appropriate models for the use of PLS in MMF is necessary. Such a transition was introduced by Winzer and colleagues for the first time. Achievable security rates in MMF links in the presence of (multiple) eavesdroppers were explored by mathematical models [33–35]. Winzer's group has experimentally shown using an SMF data link that secrecy can be achieved by information scrambling (polarization-division-multiplexed 16-ary quadrature amplitude modulation) but without using an actual MMF or a physically tapped eavesdropper [36]. Therefore, the following contributions are brought to the study of PLS in MMF within this work:

- We present a suitable diagonalization procedure for long MMFs using common-path holography and SVD (Programmable channel diagonalization for MMFs through SVD section).
- We demonstrate data transmission through MMF in which Alice and Bob exchange data by optical prescrambling, i.e., SVD diagonalization. Using wiretap codes, the link is made secure, although Eve has physical access to the channel (Experimental demonstration of confidential data links using PLS section).

- To the best of the authors' knowledge, this is the first report of a real data transmission over a physically tapped MMF, where the data have been encoded by a wiretap code and the received signal is measured at both Bob and Eve.
- We introduce a model tailored to the MMF channel to determine the secrecy rate $R_s$ based on measured TMs from the aforementioned experiment (Calculation of the achievable secrecy rate section).

## Programmable channel diagonalization for MMFs through SVD

For fiber-based multiple-input multiple-output using MMF, channel diagonalization through SVD, as shown in Fig. 3, was considered by Ho and Kahn [37] in 2013. This figure is well established for multiantenna systems [38,39]. This includes the area of PLS, where it can be the optimal strategy [40]. However, its feasibility for MMF has, to the best of our knowledge, never been shown experimentally yet. Although TM diagonalization schemes have been employed for equalization of modal XT in MMFs [25,41], it has not yet been shown to generate and prove confidential data exchange through MMF. For PLS, the hurdle is the correct measurement of the TM with increasing MMF length. During the entire TM measurement, correct phase relations on the observing position are crucial for viable light control through the MMF. Usually, the TM is measured in a specific plane on Bob's side, which is determined by the position of his camera sensor. In this case, correct phase relations imply that in each iteration, the phase is measured relative to the absolute reference point, i.e., Bob's camera. The challenge is that phase drifts between object and reference paths occur, which are induced by environmental influences such as mechanical stress and temperature fluctuations, especially when the reference is provided by a separate SMF. Therefore, the absolute position of the reference point tends to fluctuate between each measurement and, thus, between each line of the TM. This is why a reference measurement monitoring the drift of the camera position is usually involved within TM measurements [42–44]. In communication applications, increased fiber lengths are desired, because of which phase fluctuations increase in frequency. Thus, phase monitoring becomes more challenging as the frame rates of available SLMs and cameras are limited. This poses an obstacle for TM measurements with long MMFs. In contrast, guide star techniques are considerably more robust against phase drifts. With digital optical phase conjugation, the correct phase position between two playbacks is not relevant and setups of up to 1-km-long MMF links are straightforwardly implementable [45].

Here, we present an adaption of a common-path interferometer [46,47] tailored to MMF, which exhibit a dynamically changing spatial frequency spectrum within the TM measurement procedure. Using an adaptive optical pinhole realized by a digital micromirror device (DMD), we can also exploit the phase stability of common-path systems but ensure a uniform illumination profile of the reference beam in a holographic off-axis configuration. First, the light field at the receiver is split into object and reference path. While the near field is imaged onto a camera being the object wave, the far field is imaged to the surface of a DMD using a lens. Here, the brightest pixel is selected realizing spatial low-pass filtering, which is an adaptive

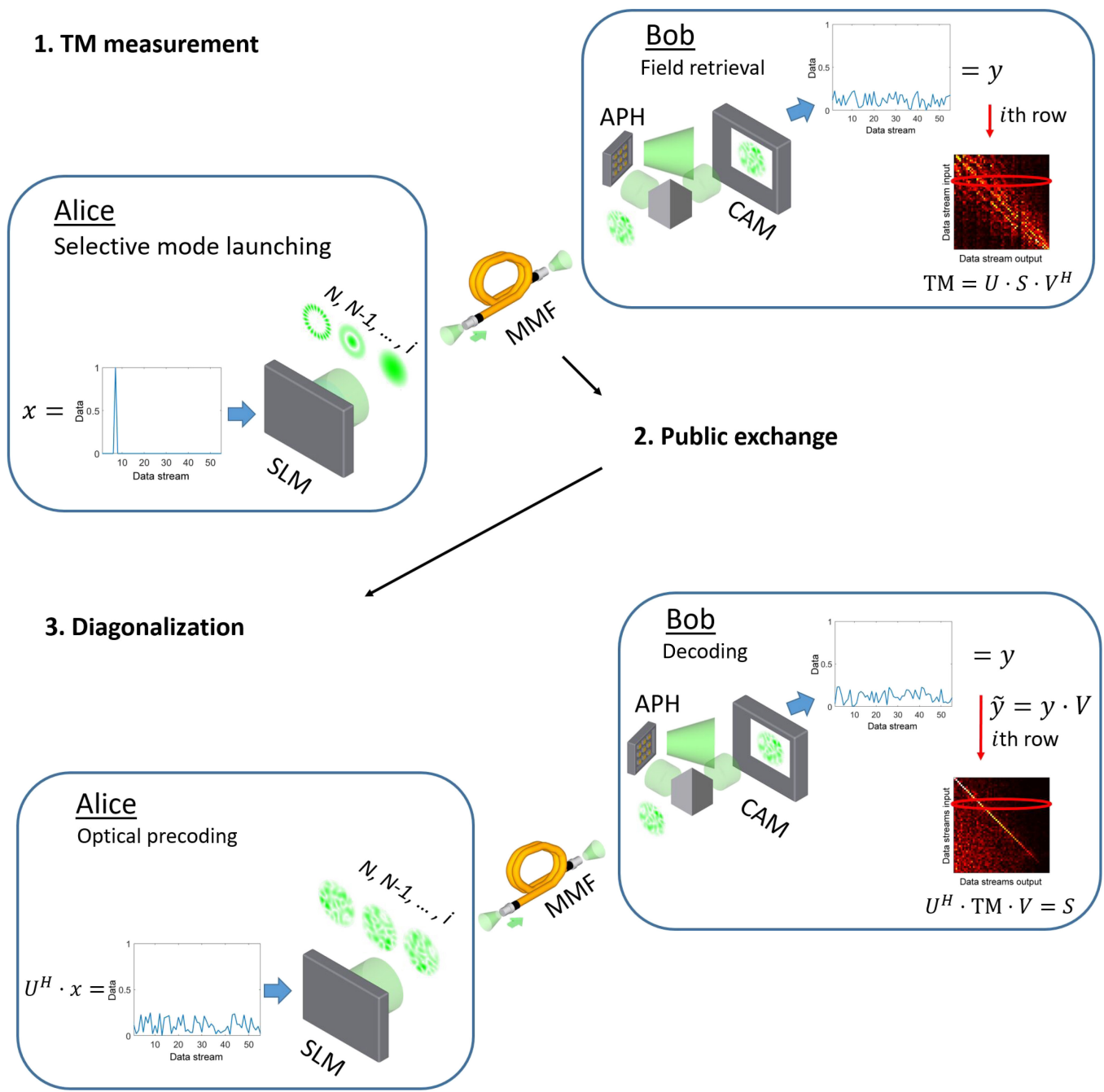

**1. TM measurement**

**Bob**
Field retrieval

APH

CAM

$= y$

$i$th row

$\text{TM} = U \cdot S \cdot V^H$

**Alice**
Selective mode launching

$x =$

$N, N-1, ..., i$

SLM

MMF

**2. Public exchange**

**3. Diagonalization**

**Bob**
Decoding

APH

CAM

$= y$

$\tilde{y} = y \cdot V$

$i$th row

$U^H \cdot \text{TM} \cdot V = S$

**Alice**
Optical precoding

$U^H \cdot x =$

$N, N-1, ..., i$

SLM

MMF

**Fig. 3.** Procedure of SVD-based channel diagonalization in MMF. During a calibration phase, Alice and Bob are determining the TM of the MMF under test. Alice uses an SLM enabling mode-selective excitation iterating through each available mode. Bob measures the scrambled light field on the receiver side, e.g., using an adaptive pinhole (APH) performing common-path holography. Through mode decomposition, Bob identifies the complex entries of the TM. After exchanging their result and applying SVD, Alice and Bob calculate matrices $U^H$ and $V$, respectively. Alice applies matrix $U^H$ for optical prescrambling using the SLM, while Bob applies $V$ for information retrieval diagonalizing the MMF channel.

pinhole. With the filtered reference wave, we superimpose the object wave to generate an off-axis hologram. During the measurement of the TM, many different DMD pixels are selected as the position of the pinhole, resulting in systematic errors in the hologram evaluation. These can be corrected straightforwardly by a look-up table. For the measurement of the TM, it is not only crucial that mode decomposition provides correct phase relations between the modes but also that the phase among the rows of the TM has the correct position relative to a freely selectable but absolute reference point. In conventional off-axis interferometers, this is achieved by a static position of an external reference wave. In the system introduced, there is a reference position that changes dynamically during the measurement process by switching different DMD pixels. Thus, the phase relation to a reference mode must additionally be measured. For this purpose, in addition to the $i$th mode whose corresponding TM row is measured, the resulting field from launching both the $i$th mode and a reference mode is measured. In our case, we choose the fundamental mode as the reference mode. It is important that the choice of the reference mode remains constant during the whole TM measurement. To determine the phase between both modes, the reconstructed fields from launching the two modes individually are digitally superimposed, and the phase difference between them is shifted until the result matches the measured field when both modes are simultaneously launched. This phase value indicates the phase position between the $i$th mode and the reference mode and is used as an offset for the $i$th TM row. This setup introduces works as a self-referencing plug'n'play solution for holographic TM measurements for long MMF links. Further information is provided in the Supplementary Materials. We show measurements on 55-mode MMF with step-index profile at up to 100 m in length.

After the TM has been measured (see Fig. 3), Alice and Bob can determine the left- and right-hand singular vectors by applying SVD to the TM: $\mathrm{TM} = U \cdot S \cdot V^H$. These matrices are unitary, which allows both participants to simply multiply them on the transmitter or the receiver side. Alice applies $U^H$ to determine an optical prescrambling, while Bob applies $V$ to the detected mode vector to retrieve information diagonalizing the channel: $U^H \cdot \mathrm{TM} \cdot V = U^H \cdot U \cdot S \cdot V^H \cdot V = S$. The remaining matrix $S$ is a diagonal matrix that has the singular values of the TM on its diagonal in descending order. Each singular value represents an available data stream the strength of which is determined by its value. Thus, by multiplying $U^H$ to any message $x$ that she wants to transmit, Alice determines the mode combination that should be launched using her SLM. The mode combination appears arbitrary, but it is well defined and matches the Bob's information retrieval $\tilde{y} = y \cdot V$. In the following section, we show an application of the SVD-based diagonalization to enable confidential data streams in an eavesdropping scenario.

## Experimental demonstration of confidential data links using PLS

To show that a confidential link can be achieved for MMF channels using PLS, several experiments have been performed. The logo of TU Dresden (TUD) serves as binary data example transmitted in all investigations, as shown in Fig. 4. In all experiments, SVD-based diagonalization is applied as channel equalizer that Alice and Bob use to transmit a message through the MMF. In our scheme, Eve gains access to the legitimate channel by physically tapping to it. Usually, Alice and Bob could expose Eve's tapping as the received intensity level on Bob's side would fluctuate dramatically. However, to evaluate the worst-case scenario from a communication perspective, the TMs of the two channels (Alice to Bob, $\mathrm{TM}_{AB}$, and Alice to Eve, $\mathrm{TM}_{AE}$) are measured using the same reception conditions for both Bob and Eve including signal-to-noise ratio (SNR) or ambient influences, e.g., mechanical stress or temperature. Therefore, we took 10-m-long MMFs and realized Eve's access via 50:50 coupling, where Bob and Eve, share the same amount of optical power on average. In the following, we show experimental results on PLS in an experimental environment. First, we apply channel diagonalization using SVD under Eve's presence and demonstrate the impact that both XT and MDL have on Eve's detection already without additional channel codes. Further, we calculate the achievable secrecy rate $R_S$ based on the measured TMs from the experiment. The results allow a quantification of the quality of the optical implementation of the SVD-based diagonalization and indicate that information-theoretic security is possible for real MMF wiretap channels. Finally, we apply wiretap codes in the experiment and showcase their power.

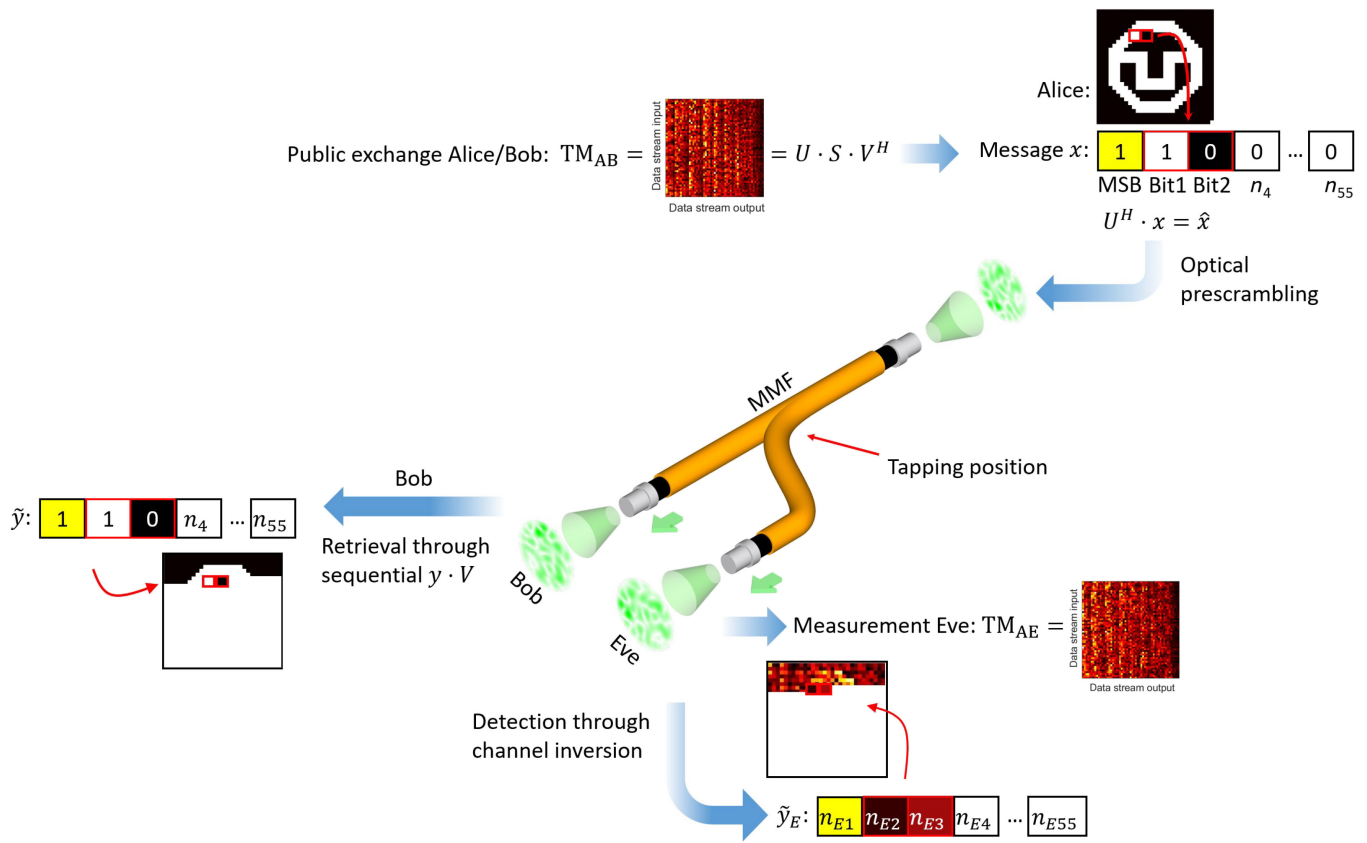

**Fig. 4.** Procedure of PLS within a real data transmission through MMF. After diagonalizing the channel using SVD, data sequences can be transmitted over the $K$ most favorable data streams represented by the $K$ highest singular values. In our case, $K=3$, and the first bit serves as most significant bit (MSB). While Alice and Bob diagonalize their channel using the SLM and SVD based on $\mathrm{TM}_{AB}$ to retrieve $\tilde{y}$, Eve measures her own TM (i.e., $\mathrm{TM}_{AE}$) and performs channel inversion. Afterward, Eve receives the tapped signal $\tilde{y}_E$. If the principle of PLS was successful, then $\tilde{y}$ contains elements of the transmitted message $x$, while, at the same time, $\tilde{y}_E$ contains noise.

### Data transmission through MMF using SVD without channel codes
In the eavesdropping experiment on real MMF, the two channel matrices $TM_{AB}$ and $TM_{AE}$ are measured simultaneously. We assume that Alice and Bob exchange their measurement result on $TM_{AB}$ over a public channel that is why it is known to Eve. Afterward, Alice and Bob apply SVD to $TM_{AB}$ and determine the matrices $U^H$ and $V$ for optical prescrambling and information retrieval. Using the $K = 3$ highest singular values, Alice transmits a binary sequence according to a $60 \times 60$-pixel image of the TUD logo, as shown in Fig. 5A. Note that the optimal allocation of $K$ can vary depending on the experiment, as explained in the Calculation of the achievable secrecy rate section. Alice uses simple on–off keying. The highest singular value serves as most significant bit and singular values 2 and 3 as actual data streams, respectively. Thus, to transmit all 3,600 pixels, Alice needs 1,800 channel uses as two data streams are used in parallel. In Fig. 5, results for data transmission to both Bob and Eve are shown. Bob can observe the transmitted image almost perfectly. The striped background in Fig. 5B is due to the repeated transmission sequence of data streams 1 to 3. The data streams 2 and 3 that carry actual data correspond to different singular values that result in different levels of the respective detected signals. Similarly, Eve can observe her received image (see Fig. 5C) with channel inversion, which is already a much distorted version of the original image due to noise amplification in the inversion process [48]. This effect could be further amplified by for instance transmitting artificial noise [49] or by physical effects enhancing MDL such as bending. However, it can be seen that some information, e.g., the outer shape of the logo, is preserved. Therefore, the transmission is not fully information-theoretic secure and only relies on the mismatch between the SVD diagonalization and Eve's channel.

### Calculation of the achievable secrecy rate
From the discussion above and the results shown in Fig. 5, we have seen that our optical implementation of the SVD

prescrambling enables a robust data transmission to Bob. However, it can also be seen that Eve still receives some information about the original transmitted image. Therefore, we need to apply techniques from the area of PLS to achieve full information-theoretic security. One of such techniques is wiretap codes that can be used as channel codes to protect the secret information [30]. The secrecy rate $R_S$ is the maximum rate of these codes at which both reliability to the legitimate receiver and secrecy against an eavesdropper can be achieved. While these results only hold asymptotically for an infinite block length of the wiretap code, it provides an upper bound on the secrecy performance of the communication system.

Therefore, we calculate the achievable secrecy rate $R_S$ for the measured MMF channels to quantify the quality of our implemented optical prescrambling at Alice's side. In particular, we compare the experimental diagonalization with a digital diagonalization, which serves as the theoretical optimum. This benchmark reveals the suitability of the optical implementation using an SLM. For our investigations, we use the measured TMs from the previous experiment.

For calculation of the secrecy rate $R_S$, we consider the following transmission scheme. Alice uses binary phase-shift keying at the input, which automatically fulfills the peak-power constraint. The constant signal power constraint is set to 10 dB. While the analysis is not limited to this particular transmission scheme, it is considered since it reflects the realistic properties of the transmission using a laser with two different output states. With the optical prescrambling using SVD discussed in the Programmable channel diagonalization for MMFs through SVD section, Alice and Bob diagonalize their channel using the TM that they have measured, which is $TM_{AB}$. This allows to transmit $K$ data streams in parallel using the subset of the $n$ available spatial degrees of freedom determined by the number of modes the MMF supports. Since the prescrambling matches only to Bob's channel, Eve needs to perform a channel inversion to equalize her channel. A detailed description of the proposed model and transmission strategy can be found in Section S3. The specification of $R_S$ determines how many bits can be transmitted with information-theoretic security. The required mathematical background is provided in Section S4. We obtain the results presented in Fig. 6. The blue curve shows the secrecy rate for the measured MMF channels where prescrambling by Alice is performed within a simulation, i.e., a perfect implementation, and therefore represents an upper bound on the performance. To evaluate $R_S$ for real data transmission through the MMF, we implemented an optical prescrambling using the SLM as shown in Figs. 3 and 4. The results on $R_S$ for the measured channels with optical prescrambling are indicated by the yellow curve in Fig. 6. These results are obtained by MC simulations with $10^5$ samples [50]. For comparison, we show the achievable secrecy rate when no diagonalization is performed at Alice as the red curve in Fig. 6. In this case, the $K$ data streams are transmitted simply over the first $K$ modes of the MMF and both receivers employ channel inversion upon reception.

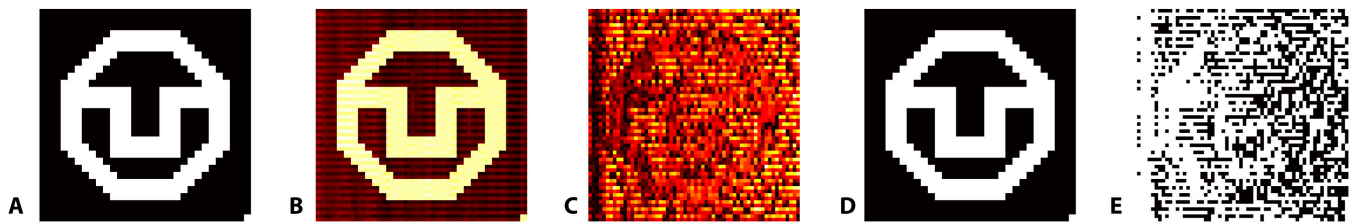

**Fig. 5.** Results on data transmission through a diagonalized MMF channel using SVD. In (A), the original image is shown. (B and C) Raw amplitude data of the received messages on Bob's and on Eve's side, respectively. These images result from the experiment introduced in Fig. 4 in which Eve gets access to 50% of the total power. The striping in (B) results from the repeated sequence on data streams 2 and 3 corresponding to different singular values. The images shown in (D) and (E) are binarized versions of (B) and (C), respectively, generated by thresholding.

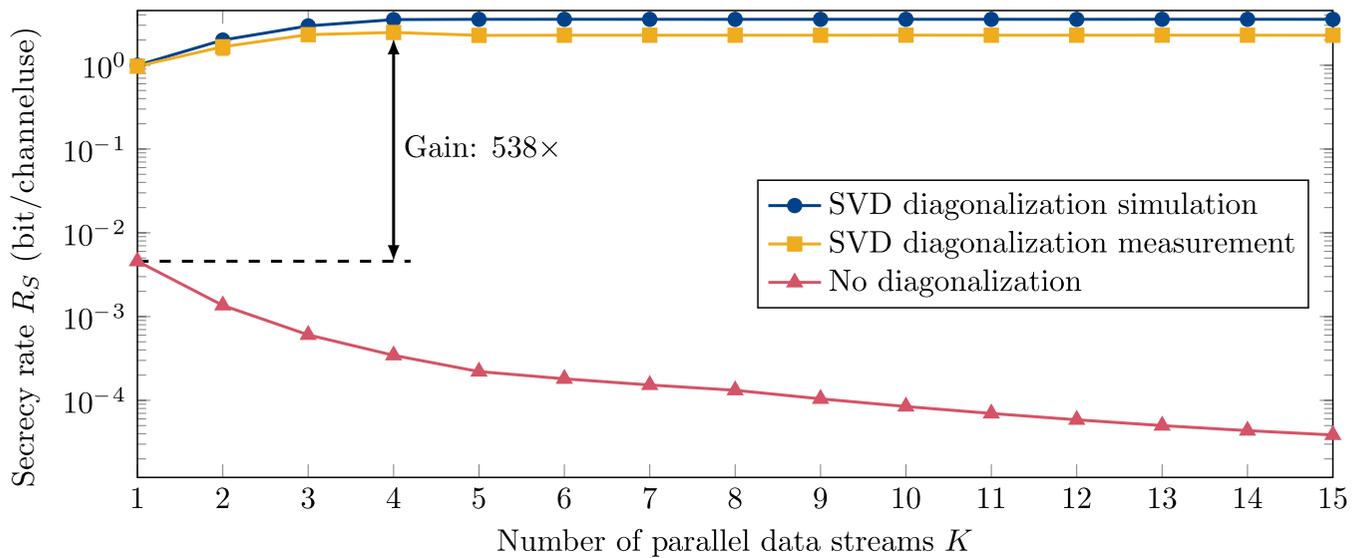

**Fig. 6.** Achievable secrecy rate $R_S$ of the MMF channels for the described binary phase-shift keying transmission scheme over the number of used data streams $K$. The first curve gives a theoretical upper bound, if SVD diagonalization could be implemented perfectly. The second curve is derived from the measurements with actual optical prescrambling. The third curve provides a comparison to the scenario without any prescrambling on the transmitter side. The indicated gap is computed between the maximum of the second curve and the maximum of the third curve.

First, it can be seen that a positive secrecy rate is indeed achievable with the considered transmission scheme. The secrecy rate for SVD diagonalization within a simulation increases with the number of data streams that are used for data transmission until it saturates. Thus, there is a minimum number of data streams on which data should be sent in parallel achieving the highest possible secrecy rate $R_S$. A parallel transmission on more data streams has no further advantage. The reason behind this is that the power for the individual data streams is allocated corresponding to the singular values of $TM_{AB}$. At high $K$, data streams with very small singular values are added, and their assigned power is close to zero. Therefore, the contribution to the secrecy rate is negligible. On the other hand, we find that there is an optimum $K$ for the actual implemented system that uses a SLM for prescrambling. In Fig. 3 and in Fig. S1A, it can be seen that the diagonalized channel between Alice and Bob is not a perfect diagonal matrix. Recall that this should in theory be a diagonal matrix containing the singular values of $TM_{AB}$ on the main diagonal, as shown in Fig. 3. However, in the practical system, the diagonalization worked properly for only a few data streams. In Fig. S1B, it is shown that within the first 10 data streams, we achieve an SNR of up to 26 dB. This result corresponds to the channel estimations of other SDM systems that operate with a single MMF core over 45 spatial modes [51]. Although some of the performance is lost because of the experimental implementation in a real system, the secrecy rate derived from the practical system is close to the upper bound. Thus, our investigations demonstrate that channel diagonalization using an SLM and SVD provides a feasible implementation for PLS in an MMF communication channel.

In contrast, there is a considerable gap of a factor of 538 from SVD diagonalization to the case where no prescrambling is used at the transmitter. Recall that data are transmitted directly on the first $K$ modes for this particular scenario. In this case, it can be noted that the secrecy rate decreases with an increasing number of data streams. This is due to MDL to the coupled eavesdropper, as the fundamental mode is the one with the least degeneracy and thus less likely to couple. Therefore, when only using this mode for data transmission, Eve gets the least information, and the highest secrecy rate is achieved. When adding more modes for data transmission, the power needs to be distributed over the modes, and more information is leaked to the eavesdropper.

### Applying wiretap codes to secure the MMF channel
PLS methods provide a powerful toolbox that enable further degradation of the received information on Eve's side while improving those on Bob's side. In our investigations, the measured TMs and the calculated secrecy rate $R_S$ are used to generate wiretap codes that enable information-theoretic security in the MMF channel [30]. Wiretap codes are a special class of channel codes, which are designed to allow a data transmission that is simultaneously reliable to a legitimate receiver and secure against an eavesdropper [22]. In this work, we apply polar wiretap codes [52] to demonstrate that secure transmissions over wiretapped MMF channels, i.e., under Eve's presence, are possible. For the transmission, we encode the TUD logo by polar wiretap codes with different secrecy rates. These binary codewords are then transmitted over the diagonalized MMF and decoded at both Bob and Eve. The prescrambling for SVD diagonalization is optically implemented by an SLM as described in the previous section. In our case, the $K = 3$ first data streams corresponding to the $K = 3$ highest singular values are again used to sequentially transmit the codewords. From the measured signals at both Bob and Eve, the resulting bit error rates (BERs) after decoding are determined. A schematic illustration of the overall system model can be found in Fig. 8, and a detailed description of the methodology is given in the Measurement of the BER using wiretap codes through MMF section.

Examples of decoded images at Bob and Eve with a BER of around 7% and 52% are shown in Fig. 7A and B, respectively. As can be observed, Bob receives an almost error-free reconstruction of the original image, while Eve receives noise. It should be emphasized that these results were obtained by an actual data transmission over a physically wiretapped MMF with a length of 10 m.

Depending on the value of $R_S$, there is a variation of the BER on both Bob and Eve's side observable. The results for the BER versus the secrecy rate $R_S$ are shown in Fig. 7C. For low values of $R_S$, it can be observed that Bob has an error-free decoding, while Eve already has a BER of around 50%. Note that this is the maximum BER for a binary transmission, since Eve effectively guesses the received bits. If $R_S$ increases, then BER on Bob's side also increases and finally also reaches 50%, as expected. A detailed list of the measurement values and experimental parameters is given in Table in the Methods section.

## Discussion

The use of PLS in MMF communication systems enables to create confidential data connections employing classic light. As PLS exploits complex phenomena of MMFs, it can be applied in high-capacity SDM networks of the future and is based on infrastructure as of today. This opens a new perspective of the already existing opportunities for information-theoretic secure exchange of sensitive data in fiber optic communication. Our results showcase that PLS is a serious alternative to other approaches providing information security.

To the best of the authors' knowledge, this work is the first to report a real data transmission over a physically wiretapped MMF channel using PLS techniques.

In our experiments, we have shown that wiretap codes can be used to exploit the asymmetry between distributed recipients on the MMF channel. As a result, data transmission from Alice to Bob is error-free, while information received on Eve's side is corrupted. This observation is reflected in the corresponding BERs: While Bob's BER is zero, Eve's BER is around 50%, which is the best possible value for a binary data transmission. We have shown in real experiments that with PLS, information can be transmitted securely with a secrecy rate of up to 3 bits per channel use. However, it should be noted that the proof of information-theoretic security according to Wyner applies to codeword lengths approaching infinity [22]. For finite block lengths, there exists a trade-off between block length, rate, reliability, and secrecy [53]. In our experiments, we transmitted codewords with finite length. Thus, although information-theoretic security is achievable in theory, current practical systems still necessitate to locally measure, for instance, the BER or another figure of merit to specify the secrecy level. Furthermore, the achievable secrecy is determined for the corresponding implementation and might vary when switching the fiber or the environment. To overcome these hurdles, more research in this area is required in the future.

We created worst-case scenarios, where legitimate and non-legitimate receivers have identical reception conditions. This means that Bob and Eve each share 50% of the transmitted power. In a practical implementation, an attacker would be revealed at lower power leakage levels since the power drain would be detected by the legitimate receiver. Nevertheless, as results show, the channel between Alice and Bob can be calibrated, and data can be transmitted confidentially through three parallel spatial data streams. We examined two different coupling positions. One is close to Alice and one is close to Bob. We observed no considerable differences under varying coupling positions. We attribute this to the pronounced modal XT occurring in step-index fibers, which is fully developed after short distances. A common tapping scenario considered is man-in-the-middle attack, where an attacker has access to the entire channel power. However, they are not feasible for fiber-optical data links. Such an attack, unlike in wireless communication networks, would require a disconnection of the link and thus be immediately revealed.

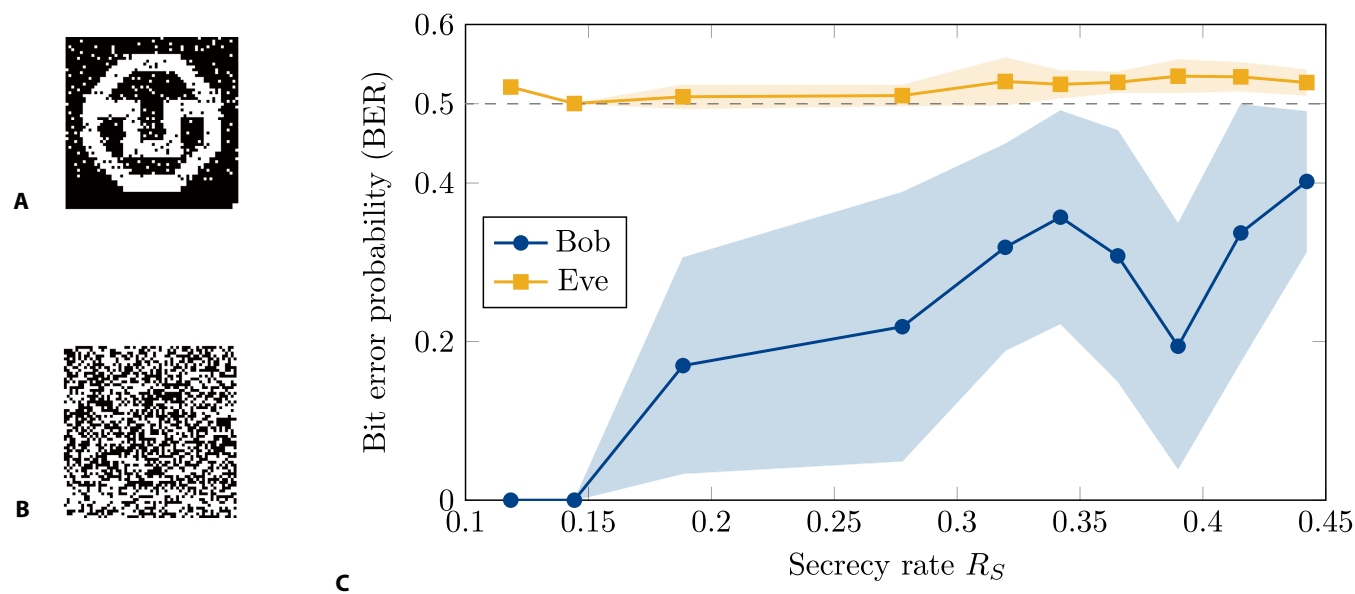

**Fig. 7.** Experimental results on data transmission through MMF using wiretap codes. Using polar wiretap codes, the confidential message, which is the TUD logo, is translated into binary codewords before transmission. The codewords are transmitted sequentially through the MMF channel, which is diagonalized using SVD. The received signals are decoded by the corresponding decoder that matches the wiretap code. The images shown in (A) and (B) are examples of the decoded image at Bob and Eve with $R_S = 0.278$, respectively. The BER of Bob's decoded image in (A) is around 7%, while Eve's BER in (B) is around 52%. In (C), the BER is evaluated for Bob and Eve under variation of the secrecy rate $R_S$. The markers indicate the mean BER values for the secrecy rates that have been tested. The filled area indicates the range of one standard deviation around the mean.

While information-theoretic security is confirmed by physical laws in quantum systems and therefore exists intrinsically, channel codes must be determined for PLS systems. Commonly, wiretap codes are one reasonable approach. For wiretap code generation, both an appropriate model and the TMs are necessary, which indicate the essential difference to QKD. Similar to QKD, PLS with wiretap codes provides information-theoretic and thereby postquantum security. The information leaked to the wiretapper does not allow to obtain the transmitted message, neither with classical nor with quantum computational resources [54]. Within the case investigated here, the model used is shown both in Methods and the Supplementary Materials, where also further details are given.

The achievable secrecy rate of PLS strongly depends on fiber parameters and the experimental apparatus for TM measurement, i.e., channel diagonalization performance. The diagonalized matrix experimentally obtained has clearly recognizable background on the secondary diagonal elements. We attribute this primarily to TM measurement in one polarization state only and thus the lack of information hidden in the orthogonal state. The complete channel information is equally distributed between two orthogonal polarization states after a short propagation distance, even if only one single polarization state is launched. This particularly applies for linear polarization states [24]. Therefore, optimized light control through MMF can only be achieved by considering both orthogonal states of polarization [45,55]. So far, this optimization step is not crucial to demonstrate the fundamental principle and utility of PLS in MMF. We use simple step-index fibers for our experiments. Compared to gradient index MMF, step-index fibers are considered less robust regarding light transport [56]. Thus, light control through them appears to be much more challenging. Therefore, we expect our results to be improved using gradient index MMF. In our investigations, a parameter configuration was used, which allows propagation of 55 modes per polarization state. The optimal number of modes for PLS incorporates a trade-off between metrological effort and information security benefits. In principle, an increasing number of modes requires a more elaborate measurement of the TM, which is why the lowest possible number of modes is advantageous as each additional mode requires at least one more measurement. In addition, it becomes more challenging to measure the TM for all modes, as more degrees of freedom are required for wavefront shaping. However, with an increasing number of modes, another TM basis such as Hadamard [57] or focal points [42] can be used reducing the experimental effort. On the other hand, an increasing number of modes is advantageous for information security. The more spatial data streams, or subchannels exist, the more data streams with high singular values to Bob could possibly be used. At long fiber lengths, phase fluctuations occur, which must be considered for interferometric TM measurements. Usually, an SMF provides the required reference externally. However, if an external reference is used, phase fluctuations between object and reference increase in frequency, the longer the distance. This effect can be reduced by, for example, a much more robust apparatus such as a common-path setup. Therefore, we built an adaptive spectral amplitude filter with which we show diagonalizations for up to 100-m step-index fiber. By upgrading TM measurement to a common-path configuration, it can be expected that there are no fundamental obstacles with regard to the achievable fiber length [42]. If desired, a more suitable wavelength can be used for long-haul transmission, and the link can be extended with a repeater or amplifier. Another required feature of the setup is a high degree of flexibility in wavefront shaping. By diagonalizing the channel, the appropriate mode combination needs to be launched as precise as possible on the transmitter-side, which can be realized best by an SLM at the moment. On the receiver side, a sufficiently high SNR is required for interferometric measurements. Basically, the required SNR is given by the quantum efficiency of the camera sensor, which is mainly influenced by the noise level of possible amplifiers or repeaters between network nodes.

For measuring the whole TM of a 55-mode MMF, 110 measurements are necessary, where 55 are required for retrieving the rows of the TM and 55 for reference enabling phase monitoring. Currently, the acquisition takes ≈35 s, as the SLM can operate at 5 Hz. However, since there is no trigger connection between camera sensors and SLM, there is high potential to further increase the system's temporal efficiency. Although the use of DMDs is extremely lossy, they achieve operation rates that are orders of magnitude higher than those of liquid crystal on silicon displays [57]. Particularly promising are piston-like DMDs whose losses are drastically reduced compared to conventional tip/tilt devices.

Other work has demonstrated comparable data transmission techniques through MMFs using time reversal [45]. The combination of classical light calibration with the transmission of single photons for QKD systems was described. Using the TM can also be considered as a promising approach to perform calibration for single-photon transmission. In our understanding, channel diagonalization using TM measurement enables, for example, the SMF-like transmission of single photons through arbitrary mode channels, which allows the implementation of QKD protocols such as BB84 in MMF [58].

## Methods

### TM measurement

On the basis of preliminary work [59], we measured all TMs in the base of the transverse modes the MMF supports. This choice provides the smallest possible TM dimension. The mode base can be derived by solving Maxwell's equations depending on the manufacturer's specifications of both MMF and laser. On Alice's side, linear-polarized modes are excited sequentially using an SLM. We use computer-generated holograms for complex light-field generation to ensure precise launching [59]. In our case, since enough SLM pixels are available, we use superpixels. We believe that other computer-generated hologram algorithms could produce similar results. Because of modal XT, speckle patterns are received on Bob's, respectively Eve's side, and imaged onto a camera. The images interfere with a reference wave in an off-axis configuration. Compared to other holographic retrieval techniques using, for example, correlation filters [60], one line of the TM can be measured single shot. After storage, these holograms are analyzed using the angular spectrum method with which amplitude and phase are retrieved. After reconstruction from both complex field components, a mode decomposition is performed to determine modal weights. The result corresponds to one row of the TM. Drifts between object and reference are monitored by a reference measurement. In our case, the fundamental mode is launched after each TM measurement step. We track the evolution of the phase in the fundamental mode, which corresponds to the global phase drift of the system. Phase values between the reference points are interpolated to correct the TM rows for the phase drift [44]. In total, we perform 55 measurements for both the TM and reference. The detailed setup is shown and described in the Supplementary Materials.

## Fabrication of tapped MMFs

Coupling for providing Eve's access to Alice and Bob's channel was realized by fiber fusion coupling (ETEK FCPW-2000 Fiber Coupler Production Workstation). We achieved a 50:50 coupling rate using a broadband light-emitting diode source. We varied tapping positions while maintaining the same coupling conditions. In terms of reproducibility, couplings were carried out on identical 10 MMF (FG025LJA, step index; Ø, 25 μm; numerical aperture, 0.1) links. From the respective tapping position, the distance to Eve's fiber facet was kept at 1.2 m.

## Information-theoretic security with wiretap codes

It should be emphasized that the definition of information-theoretic security, which we use throughout this work, is the common notion of strong secrecy. By this, a communication is information-theoretically secure, if the mutual information between transmitted message and received signal at the eavesdropper decreases to zero with increasing block length of the channel code [30, chapter 3.3]. One way to achieve this is by employing the aforementioned wiretap codes as channel codes at the transmitter [30].

The coupled MMF can be modeled on the basis of the multiple-input multiple-output wiretap channel model from wireless communications [30]. An illustration of the basic wiretap channel model can be found in Fig. 8.

Alice wants to securely transmit the message $M \in \mathbb{C}^k$ to the legitimate receiver Bob. She encodes and modulates the message to obtain the signal $X \in \mathbb{C}^n$, where $n$ is the blocklength of the wiretap code, which she then transmits over the MMF channel. Bob and Eve receive the signals $Y$ and $Z$, respectively. The transmission is called information-theoretically secure, if the mutual information between message and received signal at Eve decreases to zero with increasing block length $n$ [30, chapter 3.3], i.e., if

$$\lim_{n \to \infty} \mathbb{I}(M;Z) = 0.$$

To achieve such information-theoretic security, wiretap codes can be used as channel codes when transmitting data [30]. Typically, they have a binning structure, where the secret messages of length $k$ are mapped together with confusion messages of length $r$ to codewords of length $n$ [30]. Throughout this work, we assume binary messages and codewords. Thus, for each secret message $m \in \mathbb{F}_2^k$, there exist $2^r$ different codewords $C_m \subset \mathbb{F}_2^n$, i.e., there exists an encoding function $\gamma_m : \mathbb{F}_2^r \to C_m$ that assigns a different codeword to $m$ depending on the confusion message. The confusion messages are chosen randomly at the transmitter, and we therefore have an overall stochastic encoder $\gamma : \mathbb{F}_2^k \times \mathbb{F}_2^r \to \mathbb{F}_2^n$ for the wiretap code.

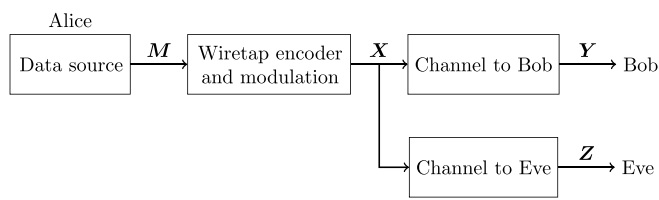

**Fig. 8.** Basic wiretap channel model with a passive eavesdropper. Alice wants to securely transmit messages $M$ to Bob. She encodes and modulates them to obtain $X$. Alice then transmits signal $X$ over the MMF channels to Bob and eavesdropper Eve, who receive signals $Y$ and $Z$, respectively.

Multiple schemes to construct wiretap codes exist, e.g., based on polar codes [52], low-density parity check codes [61], or lattice codes [62].

Additional details on the mathematical model and wiretap codes used in this work can be found in the Supplementary Materials.

## Evaluation of the achievable secrecy rate with MC simulations

The theoretical performance in terms of the achievable secrecy rate can be calculated analytically as shown in Section S4. However, this is not easily possible for the measured MMF channels with optical prescrambling due to the experimental implementation. Instead of exact calculations, we resort to MC simulations to determine the secrecy rate. On the basis of a large number of randomly generated symbols, we estimate the bit-flip probabilities of the channels to Bob and Eve and use these values for calculating the secrecy rate. The exact implementation to verify all of the presented results can be found at [50].

## Measurement of the BER using wiretap codes through MMF

The starting point for the measurements is the experimental setup shown in Fig. 4. We measured the TMs for both Bob and Eve and diagonalized the channel to Bob using the optical prescrambling with the SLM and SVD. For encoding the data, we then generate polar wiretap codes according to [52] for a given secrecy rate $R_S = k/n$. The block length is fixed to $n = 8{,}192$. Thus, we obtain an encoder for Alice that converts a confidential message of length $k$ into codewords with a defined length $n$. Note that the original image consists of 3,600 bits ($60 \times 60$ pixels), and it might therefore happen that the messages need to be split into multiple blocks of length $k$, which are then encoded individually into codewords. For the final message block, the original message bits may be padded with zeros up to the message length $k$. The codewords resulting from this process are transmitted sequentially via the $K = 3$ highest singular values using the diagonalized channel. The speckle patterns received on Bob's side are evaluated with mode decomposition, and information is retrieved according to the SVD. This principle is also shown in Fig. 3. The received bits are finally decoded by the corresponding decoder for the constructed polar wiretap code. For Eve, we have assumed that she uses the same corresponding decoder after performing a channel inversion of her channel. On the basis of the decoded signals, the number of mismatching bits for the BER is determined. We have created wiretap codes with 10 different secrecy rates, as shown in Fig. 7C. For the interval $R_S = [0.18, 0.44]$, we performed 5 measurements each. In Table, the secrecy rates, the resulting message lengths $k$, and the average BER values for both Bob and Eve are shown.

**Table.** Wiretap code parameters and measurement results (average) of the BER using a polar wiretap code with code length $n = 8{,}192$ and length of the confusion messages $r = 2$ (cf. Fig. 7).

| Message length $k$ | Secrecy rate $R_S$ | $BER_{Bob}$ | $BER_{Eve}$ |
|---|---|---|---|
| 970 | 0.12 | 0 | 0.521 |
| 1,182 | 0.14 | 0 | 0.500 |
| 1,544 | 0.19 | 0.170 | 0.509 |
| 2,275 | 0.28 | 0.219 | 0.510 |
| 2,618 | 0.32 | 0.319 | 0.528 |
| 2,802 | 0.34 | 0.357 | 0.524 |
| 2,993 | 0.36 | 0.308 | 0.527 |
| 3194 | 0.39 | 0.194 | 0.535 |
| 3,403 | 0.41 | 0.337 | 0.534 |
| 3,623 | 0.44 | 0.402 | 0.527 |

## Acknowledgments

We thank J. Peupelmann (IJP, Halsbrücke) for fabricating the tapped fibers. We also thank A. Lonnstrom and F. M. Ferreira for discussion. **Funding:** J.W.C. and E.J. acknowledge the German Research Foundation (grant nos. CZ 55/42-2 and JO 801/25-2) for funding. **Author contributions:** S.R. conducted the optical experiments with assistance from N.K., R.K., and D.K. K.-L.B. set up the model and carried out secrecy analyses. S.R. and K.-L.B. evaluated the data. J.W.C., N.K., and E.J. conceived the idea. E.J. and J.W.C. supervised the work and did project management. All authors contributed to the preparation of the paper. **Competing interests:** The authors declare that they have no competing interests.

## Data Availability

All presented calculations and MC simulations to verify the analytical results are openly published at [50]. To allow reproducing all of the results in this work, we also publish the measured TMs from which the presented results are derived.

## Supplementary Materials

Fig. S1. Investigation on SVD-based channel diagonalization in dependence of varying MMF lengths.
Fig. S2. Optical setups used for our experiments.
Fig. S3. Considered wiretap channel model with a passive eavesdropper.